\documentclass[preprint,prl,showpacs,groupedaddress,10pt,twocolumn]{revtex4}
\usepackage{amsmath}
\usepackage{amssymb}
\usepackage{graphicx}
\usepackage{graphics}
\usepackage{epsfig}

\newcommand{\ba}{\begin{eqnarray}}
\newcommand{\ea}{\end{eqnarray}}

\begin{document}

\title{Measurement of Dissipation of a Three-Level rf SQUID Qubit}
\author{Shao-Xiong Li$^{1}$, Wei Qiu$^{1}$, Zhongyuan Zhou$^{2}$, M. Matheny$%
^{2}$, Wei Chen$^{3}$, J. E. Lukens$^{3}$, and Siyuan Han$^{1}$}
\affiliation{$^{1}$Department of Physics and Astronomy, University of Kansas, 
Lawrence,
KS 66045\\
$^{2}$Department of Chemistry, University of Kansas, Lawrence, KS 66045\\
$^{3}$Department of Physics and Astronomy, Stony Brook University, Stony
Brook, NY 11794}
\date{\today }

\begin{abstract}
The dissipation-induced relaxation ($T_{1}$) time of a macroscopic quantum
system - a $\Lambda $-type three-level rf SQUID flux qubit weakly coupled to
control and readout circuitry (CRC) - is investigated via time-domain
measurement. The measured interwell relaxation time of the qubit's first
excited state, $T_{1}=3.45\pm 0.06$ $\mu $s, corresponds to an effective
damping resistance of the flux qubit $R=1.6\pm 0.1$ M$\Omega $ which is much
lower than the intrinsic quasiparticle resistance of the Josephson tunnel
junction. An analysis of the system shows that although the CRC is very
weakly coupled to the qubit it is the primary source of damping. This type
of damping can be significantly reduced by the use of more sophisticated
circuit design to allow coherent manipulation of qubit states.
\end{abstract}

\maketitle

The superconducting flux qubit, being immune to charge fluctuation, has
attracted much attention in the area of solid-state quantum computation. The
flux qubit, consisting of a superconducting loop interrupted by either one
(rf SQUID) \cite{J-R-Friedman-2000} or three Josephson junctions (persistent
current qubit or simply PC qubit) \cite{TP-Orlando-1999}, usually is a
double well potential when a magnetic flux close to one half of a flux
quantum is applied. The lowest states of each well serve as the two qubit
states corresponding to macroscopic currents circulating around the loop
with clockwise and anti-clockwise directions respectively and thus can be
discriminated by a dc SQUID magnetometer. \ Quantum superposition of
distinct macroscopic states, a prerequisite for quantum computation, was
observed in both types of flux qubits at about the same time a couple of
years ago \cite{J-R-Friedman-2000,MQC-14}. Since then rapid development in
PC qubit has led to the demonstration of coherent manipulations of a single
qubit \cite{I-Chiorescu-science}, entanglement between a PC qubit and its
readout magnetometer \cite{I-Chiorescu-nature} and between two PC qubits 
\cite{Mooij-PRL-2005}. Furthermore, the relaxation time ($T_{1}$) and
dephasing time ($T_{2}$) of the PC qubit have been measured directly using
time-domain techniques providing invaluable information about mechanisms of
and solutions to decoherence \cite%
{I-Chiorescu-science,I-Chiorescu-nature,Mooij-PRL-2005,y-yu-prl2004}. In
contrast, despite intense efforts similar measurement have not been possible
in the rf SQUID qubit due to shorter decoherence times in the particular
systems measured. It is essential to study the sources of this decoherence
to understand if it is fundamental to this type of qubit or can be overcome
by more sophisticated design. One very important element of this is the
dissipation due to the coupling of the qubit to the environment. To the best
of our knowledge the dissipation rate of the rf SQUID qubit was inferred
previously from escape probability distribution measurements which can
provide only an order of magnitude estimate of the damping resistance $R$ 
\cite{Cosmelli-1,cosmelli-reply,silvestrini-prb-2004}. Furthermore, the
results have led to more questions than answers because of the lack of an
accurate knowledge about crucial sample parameters and the use of an
effective temperature that is more than ten times of the bath temperature to
extract $R$ from the data \cite{Han-Rouse-comment,cosmelli-reply}.
Therefore, a more direct and quantitative measurement of dissipation in rf
SQUID qubits is needed in order to clarify the origin and the limit of the
decoherence in the system. In this Letter, we report results of a relaxation
time measurement of an rf SQUID qubit using time-resolved techniques. In our
experiment all parameters that enter the qubit's Hamiltonian are obtained
from independent measurements which enables us to determine the damping
resistance of the qubit with significantly improved accuracy. The result
indicates that although the inductive coupling between the qubit and its
control and readout circuits (CRC) is rather weak by conventional standards,
it nevertheless is the dominant source of dissipation inducing relaxation
from excited states. Hence, great care must be taken in the design of rf
SQUID qubits and associated CRC to limit dissipation to levels allowing
coherent manipulation of qubit states \cite{Y-Makhlin-RMP}.

The qubit used in our experiment is a variable barrier rf SQUID in which the
single Josephson junction in an ordinary rf SQUID is replaced by a low
inductance dc SQUID as shown in \textrm{Fig. 1(a)}. The two-dimensional (2D)
potential energy surface of such a variable barrier rf SQUID is \cite%
{SQUID-4}%
\begin{gather}
U\left( \phi ,\phi _{dc}\right) =\left( \Phi _{0}^{2}/L\right) \left[ \left(
\phi -\phi _{x}\right) ^{2}/2+g(\phi _{dc}-\phi _{xdc})^{2}/2\right.   \notag
\\
\left. -\beta _{0}\cos \pi \phi _{dc}\cos 2\pi \phi +\delta \beta \sin \pi
\phi _{dc}\sin 2\pi \phi \right]   \label{U}
\end{gather}%
Here, $\Phi _{0}\equiv h/2e$ is the flux quantum; $g\equiv L/2l$ is the
ratio of the inductances of the rf SQUID and dc SQUID; $\beta _{0}\equiv
2\pi LI_{c}/\Phi _{0}$, $\delta \beta \equiv 2\pi L\left(
I_{c2}-I_{c1}\right) /\Phi _{0}$, $\phi _{x}$ $\left( \phi _{xdc}\right) $
and $\phi $ $\left( \phi _{dc}\right) $ are the flux applied to and the net
flux enclosed in the rf SQUID (dc SQUID) in units of $\Phi _{0}$. A plot of
the qubit's first four energy levels is shown in \textrm{Fig. 2(a)}. Notice
that $\phi _{x}$ sets the energy bias $\varepsilon $ between the two wells
while $\phi _{xdc}$ determines the tunnel splitting $\Delta .$ Thus the
energy level structure of the qubit (e.g., $\varepsilon $ and $\Delta $) can
be varied \textit{in situ} by adjusting $\phi _{x}$ and $\phi _{xdc}$. The
qubit is inductively coupled to a hysteretic dc SQUID magnetometer and the
state of the qubit can be determined by measuring the flux-dependent
switching current of the magnetometer \cite{I-Chiorescu-science}.

\begin{figure}[ptb]
\begin{center}
\epsfig{file=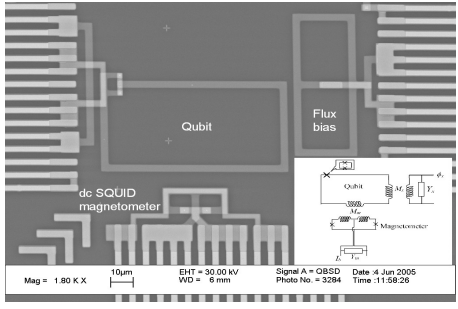,height=6cm,width=4cm,angle=-90}
\caption{Scanning microscope image of the rf SQUID qubit. Inset: Equivalent
circuit of the qubit inductively coupled to control (flux bias) and readout
(magnetometer) circuits. For $\protect\omega /2\protect\pi >5$ GHz $%
Y_{x}\approx Y_{m}\simeq R_{0}^{-1}$.}
\end{center}
\end{figure}

\begin{figure}[ptb]
\begin{center}
\includegraphics[height=6cm,width=5cm]{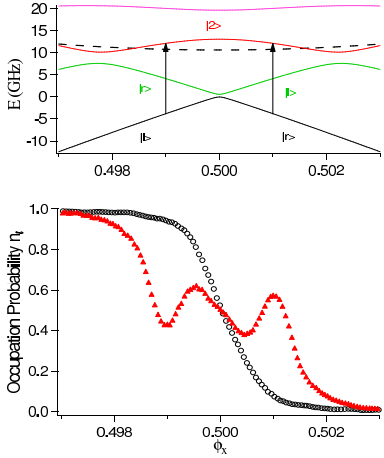}
\end{center}
\caption{(color online) (a) Enrgy of the first four eigenstates of the
qubit. Dashed line is the top of the potential barrier between the wells.
Vertical arrows indicating the position of flux bias where the level spacing
equal to $16.0$ GHz. (b) Measured occupation probability of the $|l\rangle $
state as a function of flux bias without (empty circles) and with (solid
triangles) $16.0$ GHz contonuous microwave irradiation.}
\end{figure}

In our experiment the rf SQUID is configured as a $\Lambda $-type
three-level qubit with each potential well having only one level as the $0$
and $1$ logic states and an auxiliary level just above the potential barrier
as shown in \textrm{Fig. 2(a)}. Compared to a flux qubit in the usual
two-level configuration the $\Lambda $-type three-level qubit has
significantly faster gates and much lower intrinsic error rate \cite%
{Z-Zhou-PRB2002}. Moreover, since gate operations in such a three-level
qubit are via the auxiliary level, it does not involve the tunnel splitting
energy which depends exponentially on the potential barrier \cite%
{MHS-Amin-PRB}. For these reasons the three-level flux qubit puts relatively
less-strict constraints on sample fabrication. For $\varepsilon \gg \Delta $%
, the two lowest levels of the qubit are localized, corresponding to
currents circulating in opposite directions in the qubit loop. For
convenience, we denote the state localized in the left (right) well by $%
\left\vert l\right\rangle $ ($\left\vert r\right\rangle $). The energy
difference between the $\left\vert l\right\rangle $ and $\left\vert
r\right\rangle $ states can be varied continuously by sweeping the external
flux $\phi _{x}$. Microwave radiation with frequency $\omega $ matching the
energy difference $E_{2}-E_{0}$ between the short-lived auxiliary state $%
\left\vert 2\right\rangle $ and ground state (e.g., $\left\vert
r\right\rangle $ for $\varepsilon >0$) induces transitions between them. In
general, the qubit can relax to the ground state from the auxiliary state
via two pathways: $\left\vert 2\right\rangle \rightarrow \left\vert
l\right\rangle $ $\rightarrow \left\vert r\right\rangle $ and $\left\vert
2\right\rangle $ $\rightarrow \left\vert r\right\rangle $, assuming $%
\varepsilon >0$. The probability of finding the qubit in the first excited
state depends on various transition rates (including both stimulated
absorption/emission and spontaneous emission) between the three states
involved. The lifetime of first excited state, $T_{1}$, is proportional to
damping resistance, $R=1/${Re}$[Y(\omega )]$, where $Y$ is the admittance
seen by the qubit and $\omega $ is the transition frequency of the relevant
radiative decay process \cite{AI-Larkin-JETP1986,DSP-2}. Therefore,
dissipation in an rf SQUID qubit can be investigated quantitatively via
direct measurement of $T_{1}$ between qubit states.

The sample was fabricated using a self-aligned Nb trilayer process \cite%
{SBU-trilayer}. The size of the rf SQUID loop is $50\times 100$ $\mu $m$%
^{2}. $ In order to extract the qubit damping resistance from the measured $%
T_{1}$ time one must have an accurate knowledge of key SQUID parameters,
such as total junction capacitance $C$, loop inductance $L$, and $\beta _{0}$
and $\delta \beta $. In our experiment, the loop inductance $L=205\pm 5$ pH
and $g=17.0\pm 0.6$ are estimated using a 3D inductance calculation program, 
$\beta _{0}=3.70\pm 0.02$ and $\delta \beta \leq 0.05$ were determined from
the maximum size of the measured hysteretic $\phi \left( \phi _{x}\right) $
loop (taking into account the effect of macroscopic quantum tunneling) and $%
\phi \left( \phi _{x}\right) $ at $\phi _{xdc}=1/2$, respectively \cite%
{MQT-3,SQUID-6}. The total capacitance of the junctions in the rf SQUID, $%
C=65$ $\pm 2$ fF, is obtained from microwave spectroscopy which agrees very
well with the value determined from the total size of the junctions ($1.3$ $%
\mu $m$^{2}$) and the specific capacitance of the Nb/AlO$_{\text{x}}$/Nb
trilayer ($45$ fF/$\mu $m$^{2}$). In addition, the critical current density
inferred from $\beta _{0}$, $L$, and the junction size, $J_{c}\simeq 460$
A/cm$^{2}$, agrees very well with that measured directly from co-fabricated
large junctions. The mutual inductances between the magnetometer and the
qubit is $M_{m}$ $=3.3$ pH and that between qubit and flux bias line is $%
M_{x}=1.0$ pH, respectively. The sample is mounted in a oxygen-free copper
cell thermally anchored to the mixing chamber of a dilution refrigerator. A
superconducting shield at $\sim 0.5$ K, a cryoperm shield at $4.2$ K, and a $%
\mu $-metal shield at room temperature are used to reduce flux bias
fluctuations from the ambient magnetic field. All leads to the sample cell
are filtered with electromagnetic interference filters at room temperature,
low-pass RC filters at $1.4$ K, and microwave filters at mixing chamber
temperature. A cryogenic coaxial microwave cable, with $30$ dB and $20$ dB
attenuators thermally anchored to the $1$ K plate and mixing chamber
respectively, couples microwaves to the sample. Battery-powered low-noise
preamplifiers are used to monitor the bias current and voltage of the
magnetometer. All ac-powered instruments are optically isolated from the
battery-powered circuits. Diagnostic tests using low critical current
junctions ($I_{c}$ $\sim 1$ $\mu $A) indicate that extrinsic noise is
negligible at $30$ mK.

We first measure the excitation probability, $n_{l}$, of the state $%
|l\rangle $ vs. flux bias $\phi _{x}$ at constant microwave frequency and
different $\phi _{xdc}$ to determine the energy level structure of the qubit 
\cite{J-R-Friedman-2000}. By setting the amplitude and duration of bias
current pulse properly the magnetometer either switches to finite voltage or
stays at zero voltage depending on the qubit state being $\left\vert
l\right\rangle $ or $\left\vert r\right\rangle $. The result of the
spectroscopy measurement is also used to select the values of $\phi _{x}$
and $\phi _{xdc}$ for time-resolved measurements. The result is shown in 
\textrm{Fig. 2.} For $P_{mw}=0$, when $\varepsilon $ is swept from negative
to positive there is a step-like transition from $n_{l}\approx 1$ to $%
n_{l}\approx 0$ indicating the change of ground state from $\left\vert
l\right\rangle $ to $\left\vert r\right\rangle $. When irradiated by
continuous microwave (CW) a peak (dip) in $n_{l}$ appeared at $\phi
_{x}\simeq 1/2+(-)$ $0.0010$, corresponding to microwave induced excitation
from the ground state to the second excited state. The excitation are
indicated by arrows in the energy level diagram of the qubit (\textrm{Fig 2
inset}), calculated using the independently determined sample parameters and
the full 2D potential (\ref{U}). When the value of $\phi _{xdc}$ is varied,
while keeping $\omega $ constant, the position of microwave induced peaks
shifted as expected from the calculated level diagram.

\begin{figure}[ptb]
\begin{center}
\epsfig{file=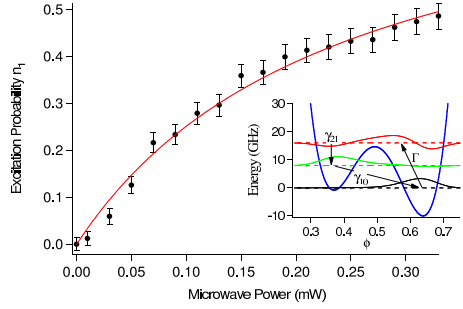,height=6cm,width=4cm,angle=-90}
\end{center}
\caption{(color online). Excitation probability of qubit vs. microwave power
at $\protect\phi _{x}=0.501$ and $\protect\omega =2\protect\pi \cdot 16.0$
GHz. The solid line is the best fit to Eq. (\protect\ref{eqn5}), where $%
P_{mw}$ is measured at the top of cryostat. Inset: The potential (projected
to the $E$-$\protect\phi $ plane), energy levels (dashed lines) and wave
functions (solid lines) of the first three eigenstates. The arrows indicate
some of the transition pathways.}
\end{figure}

Next, we measure $n_{l}$ as a function of CW power $P_{mw}$ at $\phi _{xdc}$ 
$=0.3868$ where $(E_{A}-E_{R})/\hbar =2\pi \cdot 16.0$ GHz. \textrm{Fig. 3 }%
shows the peak height as a function of the microwave power $P_{mw}$ measured
at the top of the cryostat. Since the microwave used in our experiment is
either CW or pulses with duration much longer than the decoherence time, the
dynamics of the qubit is incoherent. Hence, the time evolution of the qubit
state under microwave irradiation can be described by the following master
equation

\begin{subequations}
\begin{equation}
\partial n_{0}/\partial t=-\Gamma n_{0}+\gamma _{10}n_{1}+\left( \Gamma
+\gamma _{20}\right) n_{2},  \label{eqn2}
\end{equation}

\vspace{-0.3in}%
\begin{equation}
\partial n_{1}/\partial t=-\gamma _{10}n_{1}+\gamma _{21}n_{2},  \label{eqn3}
\end{equation}

\vspace{-0.3in}%
\begin{equation}
\partial n_{2}/\partial t=\Gamma n_{0}-\left( \Gamma +\gamma _{21}+\gamma
_{20}\right) n_{2},  \label{eqn4}
\end{equation}%
where, $n_{0}\left( t\right) $, $n_{1}\left( t\right) $, and $n_{2}\left(
t\right) $ are the normalized populations of the ground state, the first
excited state, and the second excited (the auxiliary) state, $\Gamma $ is
the rate of stimulated transitions between the ground state and the
auxiliary state, and $\gamma _{ij}$ is the spontaneous decay rate from $%
\left\vert i\right\rangle $ to $\left\vert j<i\right\rangle $ ($i,j=0,1,2$).
Note, in our experiment we have $|l=0\rangle $ ($|r=0\rangle $) for $\phi
_{x}<1/2$ ($\phi _{x}>1/2$) while $|a=2\rangle $ regardless the value of $%
\phi _{x}$. For weak microwave fields one has $\Gamma =\alpha P_{mw}$
according to Fermi's golden rule, where $\alpha $ is a coupling constant
dependent of the microwave circuit used. Under irradiation of CW, the
three-level qubit is in a steady state, $\partial n_{i=1,2,3}/\partial t=0$,
and \textrm{Eq. (2)} can be solved exactly to give 
\end{subequations}
\begin{equation}
n_{1}=\frac{P_{mw}}{\left( 1+2\gamma _{10}/\gamma _{21}\right)
P_{mw}+(\gamma _{10}/\gamma _{21})\left( \gamma _{20}+\gamma _{21}\right)
/\alpha }.  \label{eqn5}
\end{equation}%
Fitting the measured $n_{l}$ vs. $P_{mw}$ to Eq. (\ref{eqn5}) yields $\gamma
_{10}/\gamma _{21}=0.072\pm 0.040$ and $\left( \gamma _{20}+\gamma
_{21}\right) /\alpha =4.0\pm 0.8$ mW. This value of $\gamma _{10}/\gamma
_{21}$ corresponds to a ratio between the two relevant matrix elements $%
|\phi _{10}/\phi _{21}|=0.26\pm 0.05$ which agrees very well with $|\phi
_{10}/\phi _{21}|=0.254$ calculated using the independently determined qubit
parameters.

\begin{figure}[tbp]
\begin{center}
\epsfig{file=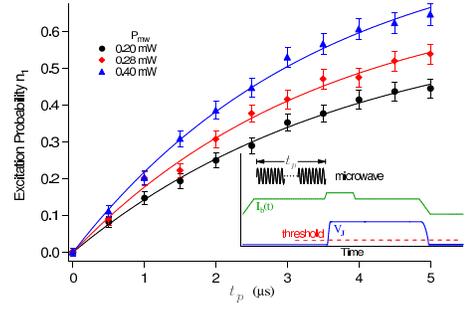,height=6cm,width=4cm,angle=-90}
\end{center}
\caption{(color online). Excitation probability of the first excited state
vs. pulse width at three different levels of microwave power. The solid
lines are best fits to Eq. (\protect\ref{eqn6}). The resulting $T_{1}$ is
independent of the microwave power applied. Inset: Signals employed in
time-domain measurement of qubit excitation probability. Top: Amplitude of
microwave applied to the qubit. Middle: Bias current of the magnetometer.
The switching current of the ground (first excited state) state is greater
(less) than the maximum bias current. Bottom: Voltage across the
magnetometer.}
\end{figure}

Although the steady state measurement provided useful information about the
system one cannot extract $T_{1}\equiv \gamma _{10}^{-1}$ from it alone. To
obtain $T_{1}$ we used a time-domain technique. We start each measurement
cycle by applying a microwave pulse of duration $t_{p}$. A bias current
pulse is then applied to the magnetometer to readout the state of qubit. The
flux bias is kept constant at $\phi _{x}=0.5010$. The time evolution of $%
n_{l}$ is obtained by varying $t_{p}$ and repeating the cycle more than $%
1000 $ times for each value of $t_{p}$. A few milliseconds of delay was
inserted between successive cycles to allow the qubit to relax to the ground
state before each new measurement cycle. \textrm{Fig. 4} shows $n_{l}(t_{p})$
for three different microwave power levels. Since $\Gamma /\left( \gamma
_{20}+\gamma _{21}\right) =P_{mw}/(4.0$ mW$)\leq 0.1$, we have $\Gamma \ll
\gamma _{20}+\gamma _{21}$. Under this condition, $n_{2}(t)$ rapidly
saturates to a constant $n_{2\infty }$, which depends on $P_{mw},$ at $%
t>t_{0}\equiv 1/\left( 2\Gamma +\gamma _{21}+\gamma _{20}\right) \ll \gamma
_{10}^{-1}$ and (\ref{eqn3}) can be solved analytically, by replacing $%
n_{2}(t)$ with $n_{2\infty }$, to give:

\begin{equation}
n_{1}\left( t\right) =n_{2\infty }(\gamma _{10}/\gamma _{21})^{-1}\left(
1-e^{-\gamma _{10}t}\right) .  \label{eqn6}
\end{equation}%
We verified the validity of the approximate solution (\ref{eqn6}) by solving 
\textrm{Eq. (2)} numerically. The result is in excellent agreement with (\ref%
{eqn6}) at $t>t_{0}$. Since $\gamma _{10}/\gamma _{21}$ is known from the
steady state measurement, fitting the measured $n_{l}(t)$ to (\ref{eqn6})
gives $T_{1}\equiv \gamma _{10}^{-1}=3.45\pm 0.06$ $\mu $s. In addition, the
results of steady state and time dependent excitation measurements together
yield $t_{0}\approx 70$ ns and $\alpha \approx 3.0$ mW$^{-1}\mu $s$^{-1}$
confirming the applicability of (\ref{eqn6}).

It is well known that the damping resistance's contribution to $T_{1}$ time
of an rf SQUID\ qubit is proportional to $R$ \cite{AI-Larkin-JETP1986}%
\ba
T_{1}^{-1}&=&\gamma _{10}=(2\pi /\hbar )(E_{1}-E_{0})(R_{Q}/R)\left\vert \phi
_{10}\right\vert ^{2}  \nonumber \\
&&\{1+\coth [(E_{1}-E_{0})/2k_{\text{B}}T]\},  
\label{eq7}
\ea
where $R_{Q}=h/4e^{2}$ is the resistance quantum. From (\ref{eq7}), using
the value of $\phi _{10}=1.0\times 10^{-2}$ calculated from the
independently determined qubit parameters, we have $R=1.6\pm 0.1$ M$\Omega $%
, which is more than $10^{2}$ times lower than the measured quasiparticle
resistance of co-fabricated junctions \cite{Vijay-2005}. Hence, we conclude
that the contribution of quasiparticles to qubit dissipation is negligible.
On the other hand, weak but finite inductive coupling to the control and
readout circuits result in a damping resistance $R\simeq \{${Re}$%
[Y_{m}\left( \omega _{10}\right) +Y_{x}\left( \omega _{10}\right) ]\}^{-1}$,
where $Y_{m}$ and $Y_{x}$ are the admittances of the magnetometer and the
flux bias circuits seen by the qubit and $\omega _{10}\equiv
(E_{1}-E_{0})/\hbar $ is the transition frequency between the $|l\rangle $
and $|r\rangle $ states, respectively. For the CRC used in our experiment
which is illustrated in \textrm{Fig. 1(b)} 
\ba
R^{-1}&\simeq &\text{{Re}}(Y_{m}+Y_{x}) \nonumber \\
&\simeq & R_{0}^{-1}[(M_{m}/L)^{2}(\Delta
L_{J}/2L_{m})^{2}+(M_{x}/L)^{2}]  \label{eq8}
\ea
around $\omega _{10}$, where $\Delta L_{J}$ is the difference of Josephson
inductances of the two junctions in the magnetometer and $L_{m}$ is the sum
of the geometric inductance ($\sim 40$ pH) and the Josephson inductances of
the magnetometer loop. Since the high frequency dampings (at $\omega \sim
\omega _{10})$ of the flux bias and magnetometer's bias/measurement circuits
are essentially the same, we model them by a shunting impedance $R_{0}$.
Substituting the values of $M_{m},$ $M_{x}$, $L,$ and Josephson inductances
of the two junction evaluated at working point of the magnetometer ($%
I_{b}=1.0$ $\mu $A, $\phi _{m}=0.45$) into (\ref{eq8}) yields $R_{0}=69\pm 5$
$\Omega ,$ which agrees very well with $R_{0}\simeq 70$ $\Omega $ derived
from the measurement of escape rate of the magnetometer in phase-diffusion
regime \cite{Mannik-2005}. The value of $R_{0}$ obtained is typical of the
high frequency impedance of transmission lines indicating that the interwell
relaxation of the qubit is mainly induced by its coupling to electromagnetic
environment through the flux bias and readout circuits. In principle,
changing the working point of the magnetometer and further reducing coupling
between the qubit and CRC could result in longer coherence time.
Unfortunately, it could not be done in our experiment since the
magnetometer's maximum critical current, $I_{c0}=9.50$ $\mu $A, is more than
three times higher than the target and consequently the flux applied to the
magnetometer ($\phi _{m}$) is bounded in a very tight range slightly below $%
\Phi _{0}/2$ to achieve nearly single-shot readout. On the other hand,
according to Eq. (\ref{eq8}), contribution to the damping resistance from
the integrated on-chip flux bias line is $1/${Re}$(Y_{x})\simeq 2.9$ M$%
\Omega $. The use of off-chip flux bias, as those employed in the
experiments observing coherent oscillations in the persistent current qubits 
\cite{I-Chiorescu-science,I-Chiorescu-nature,Mooij-PRL-2005}, would
significantly increase $1/${Re}$(Y_{x})$ but this approach would be very
difficult to apply to a circuit containing many qubits required for
practical quantum information processing. Currently, we are developing
advanced designs for the qubit bias and readout circuits that are predicted
to decrease their contributions to the qubit damping by several orders of
magnitude. In addition, due to limitations in our microwave coupling
circuit, this rf SQUID qubit was configured as a magnetometer which makes it
susceptible to ambient field fluctuations. This source of decoherence can be
greatly suppressed by using qubits with gradiometer geometries.

In summary, the dissipation of a Nb rf SQUID qubit, which is coupled
inductively to the flux bias and readout circuits, is determined by
measuring the interwell relaxation time between the ground state and first
excited state with the $\Lambda $-type three-level configuration.
Time-domain measurement of the excitation probability of the first excited
state yields $T_{1}=3.45\pm 0.06$ $\mu $s corresponding to a damping
resistance $R=1.6\pm 0.1$ M$\Omega $ for the qubit Analysis of the system
indicates that the dominant sources of qubit dissipation are the flux bias
and magnetometer readout circuits. Since this kind of dissipation-induced
qubit decoherence can be greatly suppressed with more sophisticated designs
we believe it does not impose a fundamental limit to this type of qubit.

We thank K. K. Likharev for useful discussions. The work is supported in
part by NSF (Grant No. DMR-0325551) and by AFOSR, NSA and ARDA through
DURINT (Grant No. F49620-01-1-0439).

\bibliographystyle{prsty}
\bibliography{Li-mqt}

\end{document}